%% file: PipelinePaper_jrnl.tex
\documentclass{article}

\usepackage[pdftex]{graphicx}
\DeclareGraphicsExtensions{.pdf,.jpeg,.png}
\usepackage[cmex10]{amsmath}
\usepackage[utopia]{mathdesign}
\usepackage[OMLmathrm,OMLmathbf]{isomath}
    
\begin{document}
%
\title{Large-Scale Automatic Reconstruction of Neuronal Processes from Electron Microscopy Images}
%
%
%
\renewcommand\footnotemark{}
\renewcommand\footnoterule{}
\author{Verena Kaynig \and Amelio Vazquez-Reina \and Seymour
  Knowles-Barley \and Mike~Roberts \and Thouis R. Jones \and Narayanan~Kasthuri \and
  Eric~Miller \and
  Jeff~Lichtman \and
        Hanspeter~Pfister
\thanks{V. Kaynig, M. Roberts, and H. Pfister are with the School of Engineering and Applied Sciences, Harvard University.}%
\thanks{A. Vazquez-Reina is with the School of Engineering and Applied Sciences, Harvard University and with the Department of Computer Science at Tufts University.}%
\thanks{T.R. Jones is with the School of Engineering and Applied Sciences and with the Department of Molecular and Cellular Biology, Harvard University}
\thanks{S. Knowles-Barley, N. Kasthuri, and J. Lichtman are with the Department of Molecular and Cellular Biology, Harvard University.}%
\thanks{E. Miller is with the Department of Computer Science at Tufts University.}}
%
%

\markboth{IEEE Transactions on Medical Imaging,~Vol.~1, No.~1, February~2013}%
{Shell \MakeLowercase{\textit{et al.}}: Large-Scale Automatic Reconstruction of Neuronal Processes from Electron Microscopy Images}
%



\maketitle

\begin{abstract}

Automated sample preparation and electron microscopy enables acquisition of very large image data sets. These technical advances are of special importance to the field of neuroanatomy, as 3D reconstructions of neuronal processes at the nm scale can provide new insight into the fine grained structure of the brain. Segmentation of large-scale electron microscopy data is the main bottleneck in the analysis of these data sets. 
In this paper we present a pipeline that provides state-of-the art reconstruction performance while scaling to data sets in the GB-TB range. First, we train a random forest classifier on interactive sparse user annotations. The classifier output is combined with an anisotropic smoothing prior in a Conditional Random Field framework to generate multiple segmentation hypotheses per image. These segmentations are then combined into geometrically consistent 3D objects by segmentation fusion. 
We provide qualitative and quantitative evaluation of the automatic
segmentation and demonstrate large-scale 3D reconstructions of
neuronal processes from a $\mathbf{27,000}$ $\mathbf{\mu m^3}$ volume of brain tissue over a cube of $\mathbf{30 \; \mu m}$ in each dimension corresponding to 1000 consecutive image sections.
We also introduce Mojo, a proofreading tool including semi-automated correction of merge errors based on sparse user scribbles.
\end{abstract}


%
\maketitle

\input{Introduction}

\input{relatedWork}
\section{Overview}
\input{workflow}

\input{evaluation_metrics2}
\input{data_sets}
\section{Region Segmentation}
\input{membraneClassification}
\input{results_InteractiveTraining}

\input{gapCompletion}

\input{results_membraneSegmentations}

\section{Segmentation Fusion}
\input{fusion}
\input{results_fusion}

\section{Semiautomatic Proofreading with Mojo}
\input{proofreading}

\section{Parallel Implementation}
\input{implementation_and_scalability}
\input{results_largeDataSet}

\section{Conclusions}
\input{conclusion}

\section*{Acknowledgment}
The authors would like to thank Daniel Berger for providing manual segmentation, and Nancy Aulet for interactively annotating the training data. We also would like to thank Jan Funke for providing the Sopnet results, and Bjoern Andres for providing an efficient implementation of variation of information \cite{andres:vi}. 
This work has been partially supported by NSF grants PHY 0938178, OIA 1125087, NIH grant 2R44MH088088-03, Transformative R01 NS076467, the Gatsby Charitable Trust, Nvidia, Google, and the Intel Science and Technology Center for Visual Computing. 



\bibliographystyle{IEEEtran}
\bibliography{literature}

%

%

%
%
%




\end{document}

%% file: Introduction.tex
\section{Introduction}
\begin{figure}%
\includegraphics[width=\columnwidth]{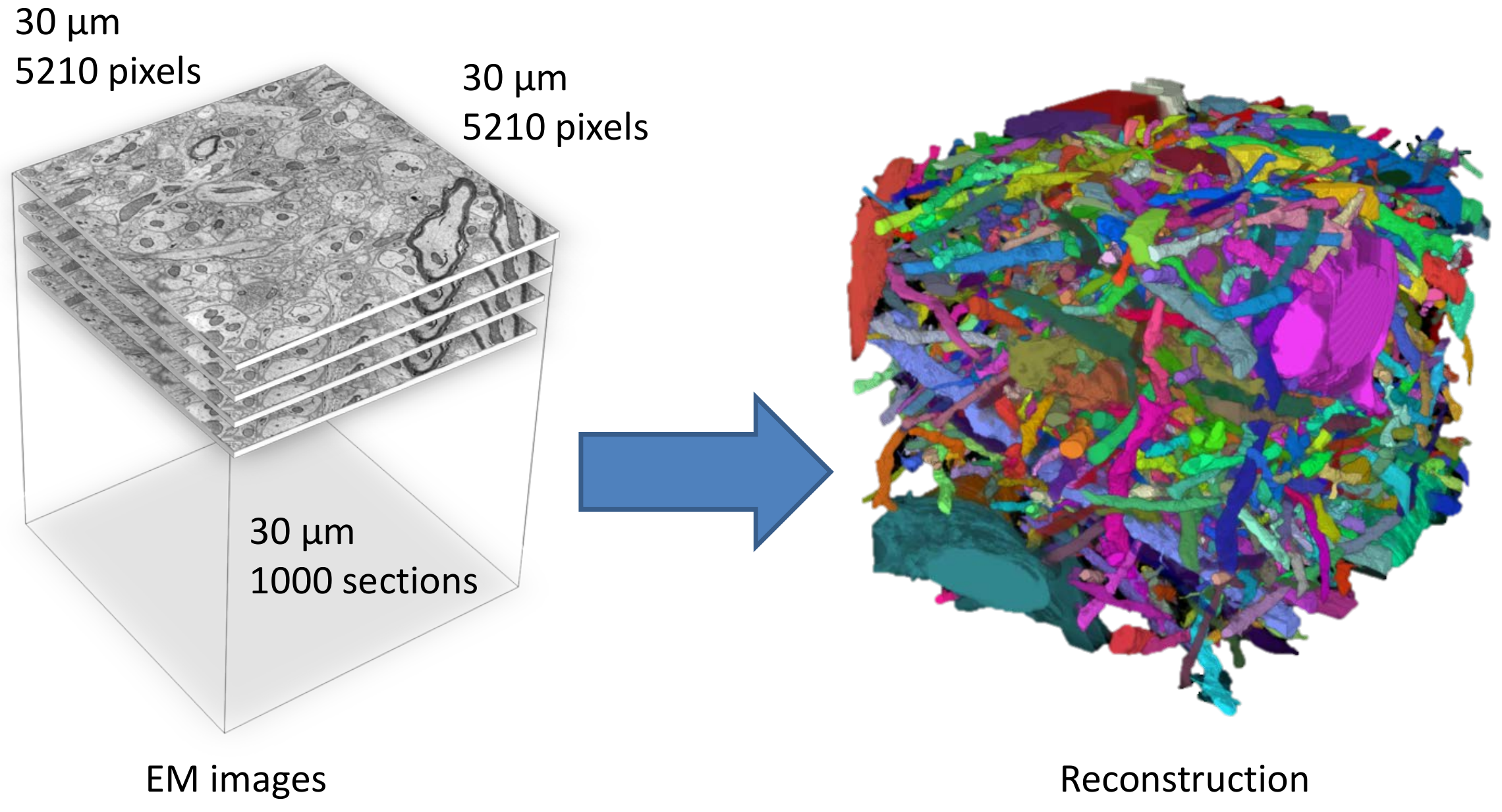}%
\caption{We propose a pipeline to automatically reconstruct neuronal processes from large-scale electron microscopy image data. The target volume consists of 1000 images with a size of $5120 \times 5120$ pixels, corresponding to $\mathrm{27,000 \; \mu m^3}$ of mammalian brain tissue. With 8 bits per pixel, the full data volume is 25 GB in size.}%
\label{fig:Introdcution_ReconstructionExample}%
\end{figure}

Brain imaging modalities such as diffusion tensor MRI or functional MRI provide important information about the brain and the connectivity between brain regions \cite{seung:12}. However, at a resolution of a cubic millimeter per voxel they provide little data about connectivity between individual neurons. Information about the anatomy and connectivity of neurons can provide new insights into the relation between the brain's structure and its function \cite{helmstaedter:12}. Such information may provide insights into the physical underpinnings of common serious disorders of brain function such as mental illnesses and learning disorders, which at present have no physical trace. Furthermore, information about the individual strength of synapses or the number of connections between two cells has important implications for computational neuroscience and theoretical analysis of neuronal networks \cite{valiant:06}. 
As the resolution of light microscopy is generally limited by diffraction, electron microscopy (EM) is a better imaging modality to resolve the brain at the level of synapses and thus provides insight into the anatomy and connectivity of neurons at nm resolution. 
To reconstruct the neuronal circuit at the level of individual cells, the field of neuroanatomy faces the challenge to acquire and analyze data volumes that cover a brain tissue volume large enough to allow meaningful analysis of circuits and detailed enough to detect synapses and thus the connectivity structure of the circuit.
Recently, significant progress has been made in the automation of sample preparation \cite{hayworth:06} and automatic image acquisition \cite{knott:08, denk:04} for electron microscopy. These techniques allow neuroscientists to acquire large datasets in the GB-TB range. 
With a resolution of $\mathrm{5 \; nm}$ per pixel,
and a section thickness of $\mathrm{50 \; nm}$, one cubic millimeter of brain tissue
results in 20,000 sections with 40 Gigapixels per image, leading to an
image volume of 800 TB. For comparison, this volume corresponds to the size of one voxel in an fMRI data set. 
With data sets this size, manual analysis is no longer feasible, leading to new challenges in automated analysis and visualization. 

In this paper we present a pipeline for semi-automated 3D reconstruction of neurons from serial section electron microscopy images. The pipeline is designed to address large data sets, while reducing user interaction to the initial training of a random forest classifier on manually annotated data and computer aided proofreading of the automatic reconstruction output. Our experiments demonstrate that the proposed pipeline yields state-of-the art reconstruction results, based on sparse annotations of only ten EM images $(1024 \times 1024 \text{ pixels})$. We provide quantitative evaluation for each step of the pipeline and an example of a reconstructed volume of $27,000 \; \mathrm{\mu m^3}$, which to our knowledge is the largest volume of conventionally stained mammalian brain tissue  reconstructed automatically (see Fig. \ref{fig:Introdcution_ReconstructionExample}).

Some of the work in this paper has been previously published \cite{kaynig:10, vazquez:11, roberts:11}. However, this is the first time we publish the complete reconstruction pipeline and and its application to large data. Specifically the novel contributions in this paper are:
\begin{itemize}
\item We demonstrate that ineractively training a random forest classifier for membrane detection not only reduces the manual annotation effort, but leads to significantly better cell region segmentations measured in terms of variation of information against manual annotated data. 
\item We combine the cell region segmentation of Kaynig et al. \cite{kaynig:10} with the segmentation fusion of Vazquez-Reina et al. \cite{vazquez:11} into a consistent pipeline leading to very good 3D reconstructions of neuronal processes from anisotropic electron microscopy data.
\item We extend the segmentation fusion approach to allow for branching structures.
\item We enable parallel processing of sub volumes via a pairwise matching scheme of segmented blocks into one consistent reconstruction volume.
\item We provide large-scale reconstruction results covering a volume of $\mathrm{27,000 \; \mu m^3}$. To our knowledge we are the first to achieve automatic reconstructions of individual spine necks in anisotropic serial section electron microscopy data prior to manual proofreading.
\item Finally, we introduce Mojo, a semi-automated proofreading tool, utilizing sparse user scribbles as described by Roberts et al. \cite{roberts:11} to correct for merge errors in the 3D reconstruction.
\end{itemize}

%% file: relatedWork.tex
\section{Related Work}

Automated reconstruction of neuronal processes has received increased attention in recent years. With electron microscopy techniques acquiring large volumes automatically, automated analysis is becoming the major bottleneck in gaining new insights into the functional structure of the brain at nm scale. The task to reconstruct the full neuroanatomy including synaptic contacts is referred to as {\it connectomics} in the literature \cite{lichtman:08}. 
A number of software packages have been developed to aid the user in manual annotation
of the images \cite{helmstaedter:11, cardona:10, fiala:05}. A complete overview of the different tools and their strength and limitations can be found in \cite{helmstaedter:12}.
In addition, semi-automatic methods have been developed to facilitate the manual segmentation process \cite{roberts:11, sommer:11, straehle:11, chklovskii:10, vazquez:09}.

In the area of fully automatic neuron reconstruction, significant improvement has been made for the segmentation of isotropic image data using a special staining method to facilitate the segmentation \cite{andres:12, andres:12b, turaga:10, andres:08, jain:07}. While these methods yield good performance for long range reconstructions, they sacrifice the staining of biologically relevant internal cell structures like vesicles or mitochondria to simplify the segmentation problem. Without staining these cell organelles, it is not possible for biologists to discriminate excitatory from inhibitory synapses or to analyze the effect of different mutations on shape and number of mitochondria. In this paper we provide long range reconstructions with conventional osmium stained images, preserving all structural information for biological analysis of the data, such as automatic mitochondria reconstruction \cite{lucchi:12, giuly:12}.

While isotropic volume data enables the direct use of 3D segmentation methods for reconstruction, the microscopy techniques for these volumes are either limited in resolution to $30 \; \mathrm{nm}$ voxels \cite{denk:04} or in the field of view to $20 \; \mathrm{\mu m^2}$ \cite{knott:08}. Serial section imaging is the only technique to record data volumes of millions of cubic micrometers \cite{bock:11}. The tissue sample is cut into ultra thin sections of $30 \; \mathrm{nm}$ and each section is imaged with an electron microscope typically at a resolution of $3-5 \; \mathrm{nm}$ per pixel. The z resolution of the resulting data volume is limited to $30 \; \mathrm{nm}$ leading to an anisotropic data volume. An interesting work by Veeraraghavan et al. aims at enhancing the z resolution by leveraging tomographic projections, but acquiring the necessary tilt images so far has not been automated for large-scale image acquisition \cite{veeraraghavan:10}.

Automatic neuron reconstruction methods for anisotropic serial section data
 typically focus on segmenting 2D neuronal regions in the high resolution images \cite{cmor:11,jurrus:11, kaynig:10} or on grouping 2D regions across multiple sections into 3D neuronal processes \cite{Funke:12, vitaladevuni:10, kaynig:10b, jurrus:08}, though some work also addresses both steps, the region segmentation and the grouping across sections together \cite{vazquez:11, chklovskii:10, mishchenko:09}. To our knowledge Chklovskii et al. describe the only pipeline so far that addresses large-scale reconstructions in the order of thousands of $\mathrm{\mu m ^3}$ \cite{chklovskii:10}. They divide the original large EM data volume into biologically relevant sub volumes of about $3000 \; \mathrm{\mu m ^3}$ which are then segmented and reconstructed. In this paper we demonstrate successful segmentation of a volume that is nine times larger
 than the result shown by Chklovskii et al. \cite{chklovskii:10}. Our experiments also demonstrate that the employed CRF framework yields better neuronal region segmentations than their use of watersheds, leading to a significant reduction in proofreading effort. 
\begin{figure*}[htp!]%
\centering
\includegraphics[width=0.8\textwidth]{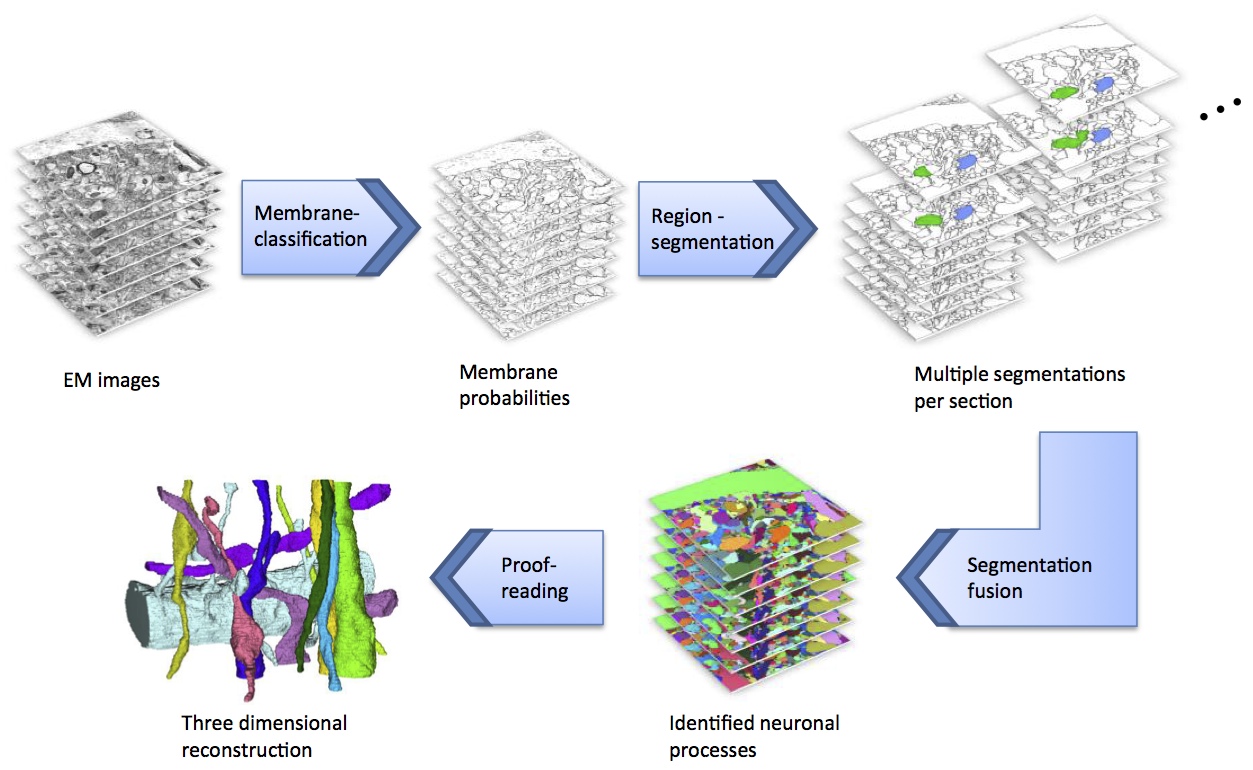}%
\caption{Illustration of our complete workflow for large-scale neuron reconstruction. First, a random forest classifier is trained by interactive sparse annotations to detect membranes in the images. Then we generate multiple region segmentation hypotheses per section. Subsequently, we find the three dimensional reconstructions as geometrically consistent segmentations across multiple sections. Finally, a manual proofreading step ensures accurate reconstructions of neuronal processes of interest.}%
\label{fig:Workflow_Pipeline}%
\end{figure*}

%% file: workflow.tex
We now provide an overview of our reconstruction workflow (see Fig
\ref{fig:Workflow_Pipeline}), as well as evaluation metrics for neuron segmentation and the data sets used for all experiments throughout this paper.

\subsection{Workflow}

For automatic segmentation methods, serial section imaging is challenging, as the resulting image data is highly anisotropic. While the xy image resolution for each section is only limited by the resolution of the microscope, the z-resolution is limited by the section thickness of about $30 \; \mathrm{nm}$. For our pipeline we assume that the image data has been previously aligned. While registration and alignment for large electron microscopy stacks is a topic of ongoing research, it is not the focus of this paper. 

Figure \ref{fig:Workflow_Pipeline} provides an illustration of the entire workflow.
The first part of the pipeline concentrates on the 2D segmentations of the high resolution section images. We first train a random forest classifier on interactive manual annotations for membrane detection. Then, we generate multiple segmentation hypothesis per section based on the classification output. Our experiments demonstrate that thresholding the membrane probability map at different intervals combined with anisotropic smoothing in a conditional random field (CRF) framework is superior to watershed segmentations of the membrane probability map \cite{kaynig:10}. We modified the original anisotropic smoothing prior to emphasize the importance of the membrane probability map over the original gray value images, leading to an improvement in segmentation performance. 

Subsequently, we leverage the previously obtained 2D segmentations and group these into geometrical consistent 3D objects using segmentation fusion \cite{vazquez:11}. This step is especially challenging for large data sets, as geometrically consistency requires context information across multiple sections. We reduced the number of features used to measure region similarity to streamline the fusion computation. In addition, we extend the original segmentation fusion model \cite{vazquez:11} to allow for the reconstruction of branching structures. We evaluate the fusion step of the pipeline and compare bipartite matchings of globally optimal groupings of sub volumes with a greedy optimization scheme. 

In the final step, the segmentation output has to be proofread by a user, to ensure correct geometries. As fully manual proofreading is labor-intensive, we introduce Mojo, a semi-automatic proofreading tool, that leverages sparse user scribbles to correct merge errors in the automatic segmentation \cite{roberts:11}.

%% file: evaluation_metrics2.tex
\subsection{Evaluation Measure}
There are two types of errors: split errors and merge errors. In 2D segmentation, a split error refers to a single region being split into two or more regions in the segmentation due to false positive cell boundary detections. A merge error is caused by a gap in the segmented cell boundaries, leading to separate regions being merged into one region in the automatic segmentation. Both errors can also occur during region grouping in 3D. Missing a branch, for example, can lead to a split error, whereas merging branches incorrectly can merge two different neural processes into one object.

As drawing of cell boundaries requires more precision than clicking on objects, it is generally faster for a user to correct split errors than merge errors. Therefore, previous work on neuron segmentation has biased the output of the automatic reconstruction towards obtaining an over-segmentation of the data \cite{chklovskii:10}. We follow a different approach in our work. Instead of biasing the pipeline towards split errors, we provide a 3D semi-automatic segmentation method in our proofreading tool to assist the user with the correction of merge errors. This allows us to focus on optimizing the overall error rate with an equal weighting of split and merge errors. We measure the quality of our segmentation by comparing it to a manual annotation using variation of information. 
The variation of information metric is a standard evaluation measure for region segmentation \cite{arbelaez:11}. It is based on information theory and compares two segmentations $S_1$ and $S_2$ based on their entropy $H$ and mutual information $I$:
\begin{equation}
VI(S_1, S_2) = H(S_1) + H(S_2) - 2I(S_1,S_2).
\label{eq:VariationOfInformation}
\end{equation}
Entropy $H$ measures the randomness of the segmentation, and mutual information $I$ measures the information that the two segmentations share. Eq. (\ref{eq:VariationOfInformation}) can be rewritten as $VI(S_1,S_2) = H(S_1|S_2) + H(S_2|S_1)$, thus, variation of information measures the amount of randomness in one segmentation given that we know about the other segmentation and vice versa. All variation of information scores reported in the paper are given in nats. 

Variation of Information can be computed efficiently and is defined for arbitrary dimensions. Thus, the same evaluation criterion can be employed to evaluate the 2D region segmentations as well as the 3D region grouping step of our pipeline. 


%% file: data_sets.tex
\subsection{Data Sets}
To demonstrate the scalability of our reconstruction workflow we use a data set consisting of 1000 sections, with $5120 \times 5120$ pixels per image. The tissue is dense mammalian neuropil from layers 4 and 5 of the S1 primary somatosensory cortex of a 5 month old healthy C45BL/6J mouse. The images were taken at a resolution of $3 \; \mathrm{nm}$ per pixel and downsampled by a factor of two, leading to a resolution in the image plane of $6 \; \mathrm{nm}$ per pixel. The section thickness is $30 \; \mathrm{nm}$. The entire data set captures a tissue volume of $30 \times 30 \times 30 \; \mathrm{\mu m ^3}$. Manual reconstruction of this data is performed by annotating the whole structure of interest. This differs from center line tracings used in previous work by Helmstaedter et al. \cite{helmstaedter:11}. Center line tracings can be annotated faster by a user than complete volume reconstructions and are sufficient to evaluate the tracing of neuronal processes over long ranges in a given volume. In contrast, our complete volume segmentations enable us to also evaluate the automatic segmentation with respect to important fine structure details, such as spine necks (see section \ref{sec:large_scale_reconstructions}), that are necessary to identify neuron connectivity.
  
In addition, we used a smaller volume consisting of 150 section images with $1024 \times 1024$ pixels per image. This volume has been densely annotated and thus captures the full range of different object sizes and variability on a small scale. Instead of carefully drawing cell boundaries, manual segmentation was performed by focusing on the regions corresponding to neuronal processes. Thus, small extracellular space between cells as well as thick or fuzzy membranes can lead to unlabeled pixels in the manual annotation. In order to preserve the duality between cell boundaries and annotated regions, we assign unlabeled pixels the label of the closest annotated region using seeded region growing. In this paper, all training and parameter validation is restricted to the first 75 images of the densely labeled data set, whereas the second half is used for testing only after all parameters of the workflow have been fixed.

%% file: membraneClassification.tex
While the texture characteristics of cell regions in electron microscopy images can vary significantly between different animal types and staining protocols, the basic appearance of the cell boundary membranes as thin, smooth, and elongated structures remains the same. Thus, instead of segmenting interior cell regions, we focus on segmenting the cell membranes to make our approach easily adaptable to a wide range of data.

\subsection{Membrane Classification}
To learn the characteristics of membranes in the electron microscopy images, we train a random forest classifier based on sparse manual membrane annotations. Random forests combine the idea of bagging decision trees with random feature selection \cite{breiman:01}. Each decision tree is built from a bootstrapped sample of the training data and at each node a random subset of the available features is selected to estimate the best split \cite{breiman:01}. For prediction, the votes of all decision trees in the forest are accumulated. As each tree can be grown and queried independently, random forests are ideal for parallelization during training and prediction, as well as in an interactive training framework. In addition, random forests are robust against over-fitting, leading to good generalization performance with few manual annotations. The parameters to tune are the number of decision trees and the size of the feature subset used to determine the best split. The performance of the classifier is not sensitive to these parameters and default values produce good results in our experiments. We employ 300 trees and we set the number of features to the square root of the total number of features, which is the default suggested by Breiman \cite{breiman:01}. To account for imbalanced training data, we follow the approach of Chen et al. \cite{chen:04}, and reduce the bootstrap sample for each tree to the size of the minority class.

The feature set extracted from the images is designed to capture the characteristics of membranes with little computational cost. Extracted features include the gray value, gradient magnitude, Hessian eigenvalues, and difference of Gaussian for the image smoothed with Gaussian filters of different kernel sizes. In addition, we convolve the image with a steerable filter at different orientations. Each filter output serves as a feature, as well as the minimal, maximal, and average output of the steerable filter for different orientations at a pixel position. 

\begin{figure}
\centering
\includegraphics[width=0.45\columnwidth]{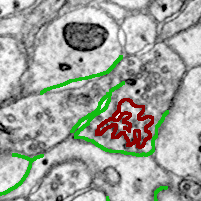}\hspace{0.25cm}
\includegraphics[width=0.45\columnwidth]{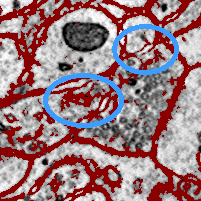}%
\caption{An example of the interactive annotation workflow. Left: An original electron microscopy image, with overlayed membrane annotations (green) and background annotations (red). Right: The thresholded membrane probability map overlayed in red. The ellipses mark a split and merge error respectively, which could be corrected by additional annotations in the next iteration.}%
\label{fig:Example_InteractiveTraining}%
\end{figure}
%
Changes in the sample preparation process or different animal types can lead to significantly different data sets, requiring a retraining of the membrane classifier. Thus, our approach aims at minimizing manual interaction. We use an interactive training approach, similar to Sommer et al. \cite{sommer:11}. The user provides sparse training annotations of membranes and the background class and interactively corrects the output of the classifier in a feedback loop. The main benefit of this method is that the annotation effort is efficiently guided towards challenging classifications and saves the user from annotating membranes that are already correctly classified. In addition, the user can bias the classification output by  accepting false positive membrane detections, for example on vesicles, as long as these do not lead to split errors in the segmented regions. Figure \ref{fig:Example_InteractiveTraining} depicts an example of the interactive annotation. 
Our experiments demonstrate that in the context of small training samples, this interactive approach outperforms complete annotation of all membranes in the images (see Figure \ref{fig:Results_InteractiveTraining}). 

%% file: results_InteractiveTraining.tex
\subsection{Interactive Training Evaluation}\label{subsec:interactiveTrainingEval}
\begin{figure}%
\includegraphics[width=\columnwidth]{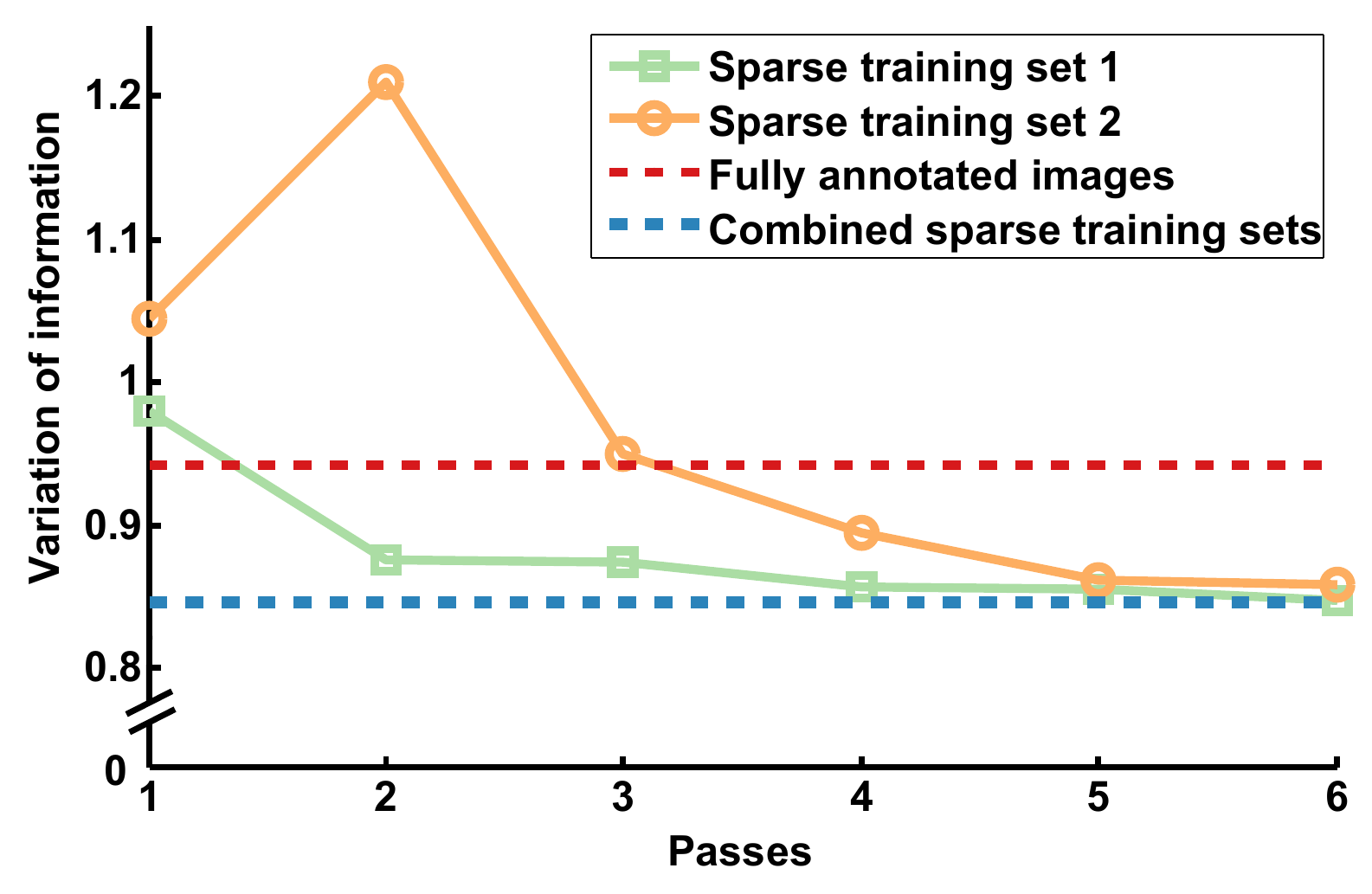}%
\caption{Evaluation of the interactive annotation approach using variation of information as the metric. Lower values correspond to better performance. After the third to fourth round of interactively correcting the classifier output (orange and green lines), the interactive approach outperforms training on fully annotated images (red). The blue line corresponds to the performance of the classifier trained on the last round of both sparse training sets together. }%
\label{fig:Results_InteractiveTraining}%
\end{figure}
To evaluate the interactive annotation against conventional training of fully annotated images, we chose a training set of five images out of the first 75 images of the 150 section data set. These images were manually selected to cover the variability in the data, from changes in image focus, contrast, or section thickness. The interactive training consists of multiple passes over the training images. Each pass consists of an annotation and a feedback phase. The user is presented with the current classification result and then manually provides additional training samples to correct misclassifications. To ensure reproducibility of the result, we repeated the interactive annotation procedure with a second data set of five training images. Figure \ref{fig:Results_InteractiveTraining} demonstrates the median performance of the interactive sparse training in terms of variation of information. The performance is measured over the validation image data, which consists of the 65 images out of 75 that were not used for training annotations. 
 
To compare the performance of sparse interactive annotations to conventional batch training on fully annotated images, an expert labeled the center of all membranes in all 10 images of both training sets. All pixels with a distance greater than five pixels from the annotated membranes are taken as background examples.
After the third to fourth pass over the images, the interactively trained classifier outperforms the classifier trained on fully annotated images.  As demonstrated by the green and orange curves in Figure \ref{fig:Results_InteractiveTraining}, the performance of the second training set is similar to the first training set.
Interestingly, the second pass on the second training set shows a significant degradation of the segmentation. In this step the annotator introduced background labels on mitochondria, leading the classifier to misclassify fuzzy membranes as background and thus to introduce gaps in the cell boundaries. In the next step these membrane misclassifications are corrected, leading to an improvement in the segmentation performance. 
The blue line corresponds to the performance of a classifier trained on the final ten training images of both interactive sparsely annotated sets. This is the classifier we use for the remaining steps of the pipeline.

%% file: gapCompletion.tex
\subsection{2D Segmentation}
%
The random forest classifier captures the main image characteristics of membranes with little manual annotation data. Previous work has shown that anisotropic smoothing of images is beneficial for the segmentation of membranes \cite{mishchenko:09}. We follow the approach of Kaynig et al. \cite{kaynig:10}, which combines the membrane probability output of the random forest classifier with an anisotropic smoothing prior for gap completion in a Conditional Random Field (CRF). 
In a CRF, the binary segmentation of all pixels as foreground or background is estimated by maximizing the a posteriori probability of the labels $y$ given the observed data $x$:
\begin{equation}
\begin{aligned}
p(y|x) \propto \exp(&\sum_{i, p \in P} \lambda_i F_{\text{state}_i}(y_p,x,p) + \\
&\sum_{j, p \in P, q \in N(p)} \lambda_j F_{\text{trans}_j}(y_p,y_q,x,p,q)).
\end{aligned}
\label{eq:CRF_MAP}
\end{equation}
$F_{\text{state}_i}$ is a state feature function of the label $y_p \in \{0, 1\}$ at pixel $p \in P$, and the image intensity values $x$, and $F_{\text{trans}_j}$ is a transition feature function of the labels $y_p$ and their neighbored labels $y_q$ in the 8-connected neighborhood $N(p)$. 
Intuitively, in our framework the state feature function estimates the probability of a single pixel as being foreground or background, whereas the transition feature function introduces dependencies between neighbored pixels, leading to smooth segmentations. 
Instead of maximizing the a posteriori probability of the labels $y$ we minimize the negative logarithm, leading to the following energy term:
\begin{equation}
\begin{aligned}
E(y) = \sum_{p \in P}E_{rf}(y_p) + & \lambda_{s} \sum_{p \in P, q \in N(p)} E_{s}(y_p,y_q) \\
 + & \lambda_{gc} \sum_{p \in P, q \in N(p)} E_{gc}(y_p,y_q).
\end{aligned}
\label{eq:graphCut_energy}
\end{equation}
The state function $F_{\text{state}_i}(y_p,x,p)$ corresponds to the data term $E_{rf}(y_p)$, which uses the output of the random forest classifier to specify the costs for label $y_p$ being membrane or non-membrane. In Eq. \ref{eq:graphCut_energy} we omit the arguments for the observed data $x$ and the pixel positions $p,q$ to simplify the equation. 

In addition, we include two smoothness terms which correspond to transition feature functions in Eq. (\ref{eq:CRF_MAP}). One is an isotropic smoothness term $ E_{s}(y_p, y_q) $, which penalizes for discontinuities in the segmentation for neighboring pixels of similar intensities. This smoothness term is widely used in graph cut approaches \cite{boykov:06}:
\begin{equation}
E_{s}(y_p,y_q) = \exp \left( - \frac{(x_p - x_q)^2}{2 \sigma_{s}^2} \right) \cdot \frac{\delta (y_p, y_q)}{\text{dist}(p,q)},
\label{eq:E_smooth}
\end{equation}
where $x_p$ is the gray value of the image at pixel $p$ and $\text{dist}(p,q)$ takes the distance between neighbored pixels into account. The Kronecker delta function $ \delta (y_p, y_q) $ equals 0 if $y_p = y_q $ and 1 otherwise. Thus, the Kronecker delta function penalizes label changes, whereas the first factor of the energy term alleviates this penalty for strong changes of contrast in the image. 

The second smoothness term $ E_{gc}(y_p, y_q) $ enhances the coliniarity of segmented pixels:
\begin{equation}
\begin{aligned}
E_{gc}(y_p, y_q) = & |<v_p,u_{pq}>| \cdot \exp \left (  - \frac{(1 - x_m)^2}{2 \sigma_{gc}^2}  \right ) \\
\cdot & \frac{\delta_{\rightarrow} (y_p, y_q)}{dist(p,q)},
\end{aligned}
\label{eq:E_gc}
\end{equation}
where $u_{pq}$ is a unit vector with the orientation of a straight line between pixels $p$ and $q$, and $v_p$ is a vector directed along the membrane. The length of $ v_p $ reflects the orientedness of the image at $p$. To measure the orientation of the membrane we use a directed filter consisting of a straight line with a thickness comparable to the membrane thickness in the training images. The term $ <v_p, u_{pq}> $ is then estimated by the response to this filter at the orientation corresponding to $u_{pq}$.
The value of $ x_m $ is the probability of pixel $x$ being a membrane, and $ \sigma_{gc}^2 $ can be estimated as the variance of these probabilities. Thus, the difference $(1 - x_m)$ weighs the energy term according to the confidence of the classifier in $x_m$ being a membrane. 
In contrast to Equation \ref{eq:E_smooth}, the factor $ \delta_{\rightarrow} (y_p, y_q) $ is not symmetric. Instead  $ \delta_{\rightarrow} (y_p, y_q) = 1$ for $y_p = 1, y_q = 0$ and  $ \delta_{\rightarrow} (y_p, y_q) = 0$ for all other cases.  This asymmetric definition ensures that $E_{gc}$ only penalizes for cuts that violate the smoothness along the direction of membrane pixels. 

The smoothness terms $E_{s}$ and $E_{gc}$ are submodular, i.e., $E(0,0) + E(1,1) \leq E(1,0) + E(0,1)$, and thus the global minimum of $E(y)$ can be efficiently found by max-flow/min-cut computation \cite{kolmogorov:04, boykov:04, boykov:06}.

For this purpose, we define a graph $G=(\mathcal{V,E})$. The set of graph nodes $\mathcal{V}$ consists of all pixels $p \in P$ and two additional terminal modes $s$ and $t$ that represent foreground and background in the segmentation. The set of directed edges $\mathcal{E}$ connects all pixels $p$ to their neighbors $q \in N(p)$. Weights are assigned to these edges as specified by the smoothness terms $E_{s}$ and $E_{gc}$. In addition, the set of edges $\mathcal{E}$ connects each pixel to two additional terminal nodes $s$ and $t$ with weights specified by $E_{rf}$. Minimizing $E(y)$ corresponds to finding the optimal cut $\mathcal{C} \subset \mathcal{E}$ such that no path exists between the terminal nodes $s$ and $t$ in the graph $G_{cut} = (\mathcal{V,E-C})$. The cut is optimal in the sense that the sum of all edge weights of all edges included in the cut is minimal.

The optimal labeling $y$ corresponds to a binary segmentation of the image into membrane and non-membrane pixels. As we are ultimately interested in the region segmentation of neuronal processes, we identify neuronal regions as connected background components and then use seeded region growing to obtain a complete tessellation of the image into segments corresponding to neuronal processes.

%% file: results_membraneSegmentations.tex
\subsection{Region Segmentation Evaluation}\label{subsec:regionsegeval}
To evaluate the performance of our 2D segmentation step we set the isotropic smoothing weight $\lambda_s=0.6$ and the anisotropic smoothing weight $\lambda_{gc} = 0.1$. The values for theses parameters were optimized over the 65 images of our validation set. A  visual example of the output is provided in Figure  
\begin{figure}%
\includegraphics[width=.49\columnwidth]{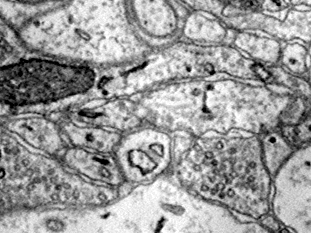}
\includegraphics[width=.49\columnwidth]{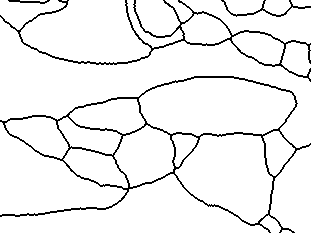}\vspace{0.25cm}\\
\includegraphics[width=.49\columnwidth]{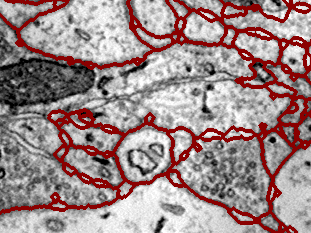}
\includegraphics[width=.49\columnwidth]{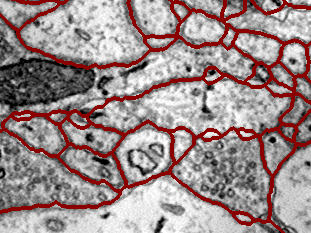}
\caption{Example segmentation of a test EM image. Top left: original image. Top right: manual annotation. Bottom left: random forest output. Bottom right: CRF segmentation. The isotropic smoothing closes small regions caused by extracellular space between cells, whereas the anisotropic term prevents shrinking bias for long elongated structures and enhances gap completion.}%
\label{fig:Results_segmentation_example}%
\end{figure}
\ref{fig:Results_segmentation_example}. 
Figure 
\begin{figure}%
\centering
\includegraphics[width=\columnwidth]{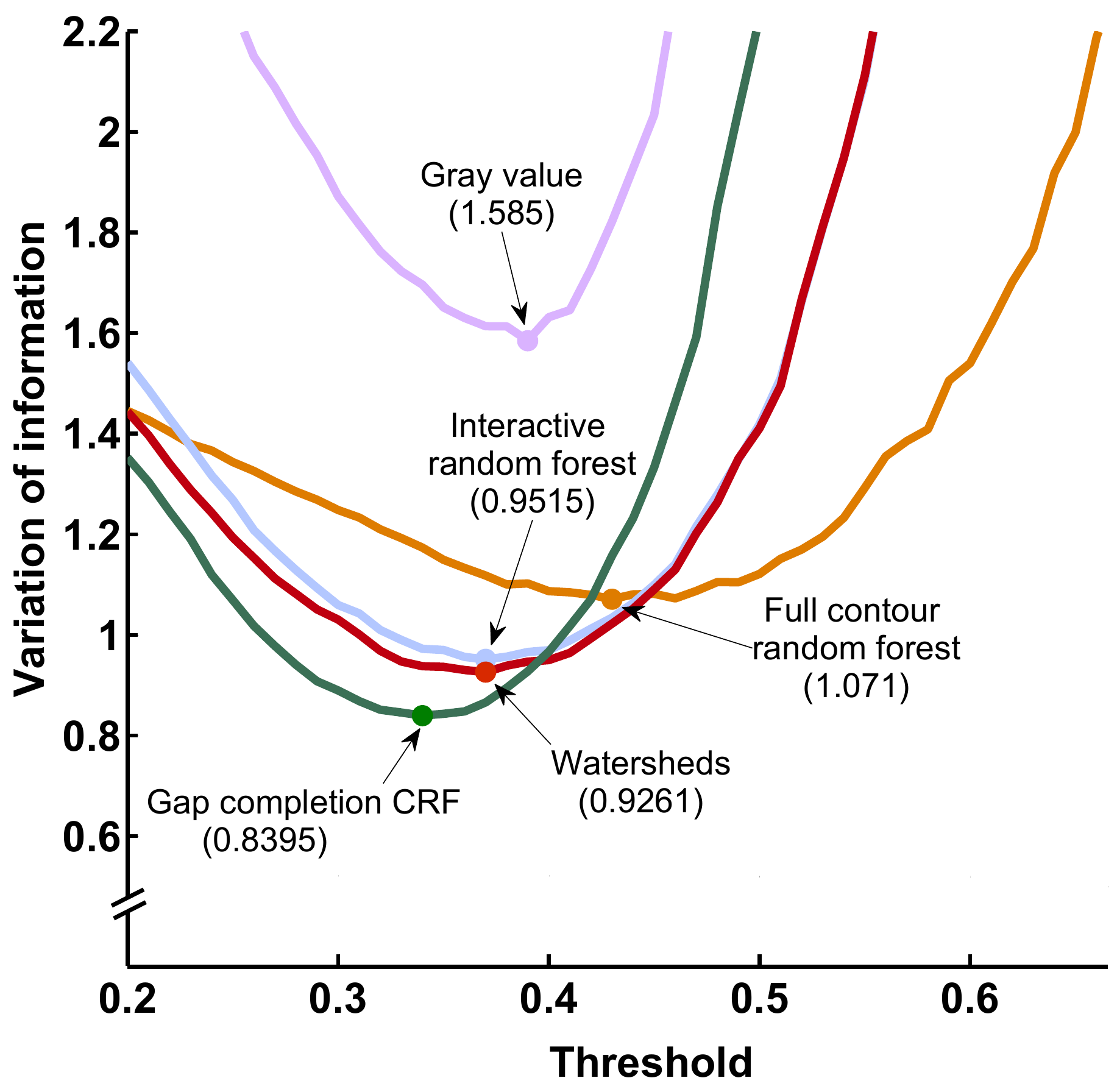}%
\caption{Evaluation of membrane segmentations on test data. The top line (violet) corresponds to the performance of thresholded gray value images. The orange line demonstrates the performance by a random forest classifier on ten fully contoured training images. The blue line shows the evaluation for a random forest trained on interactive sparse annotations. The red line corresponds to watershed segmentations of the random forest probability map. The green line corresponds to the performance of the CRF framework in our pipeline.}%
\label{fig:Results_membraneSegmentation}%
\end{figure}
\ref{fig:Results_membraneSegmentation} provides a quantitative comparison of different 2D segmentation methods over 75 test images that were not used for any parameter tuning. A random forest classifier trained on fully contoured images (orange line in Fig. \ref{fig:Results_membraneSegmentation}) gives a large improvement in segmentation performance over direct segmentation of gray values, but the random forest trained with sparse interactive annotations demonstrates a better generalization to the test data. We also compare the output of the CRF framework with gap completion against watershed segmentations, which have been widely used in previous work to generate segmentations of neuronal structures in EM images \cite{vazquez:11, straehle:11,chklovskii:10,andres:08}. While the optimal watershed segmentation performs 0.03 better in terms of variation of information than the best thresholded random forest output, the CRF framework yields an additional improvement of 0.09 over the best watershed segmentation.

%% file: fusion.tex
\subsection{Region Grouping Across Sections}\label{subsec:fusion}
The previous steps of the pipeline focus on the segmentation of neuronal processes in the 2D image plane to take advantage of the high resolution provided by the electron microscope. To extract the 3D geometry of neuronal processes, these regions need to be grouped across sections. We follow the segmentation fusion approach of Vazquez-Reina et al. \cite{vazquez:11} that allows for globally optimal groupings of regions across sections. The term fusion refers to the option to pick the best choice of geometrically consistent region groupings out of a set of possible segmentations for each section. The fusion problem is formulated as the maximum a posteriori labeling over a set of binary indicator variables. Each indicator variable $s_i$ corresponds to a possible 2D region of a neuronal process, and each indicator variable $l_j$ to a 3D link between regions of adjacent sections. If an indicator variable is activated (e.g., $s_i=1$), the correspondent region is assumed to be selected for the final segmentation, and similarly for a 3D link $l_j$. Thus, a labeling of the indicator variables corresponds to a 3D segmentation of the whole data volume.

Following the model of a CRF described in Eq(\ref{eq:CRF_MAP}) the fusion problem is modeled as:
\begin{equation}
\begin{aligned}
p(s,l|r) \propto & \exp \bigg( \sum_{i = 1}^n F_{\text{segment}}(s_i,r,i) + \\
&\sum_{j=1}^m F_{\text{link}}(l_j,r,j) \bigg) \psi(s,l) .
\end{aligned}
\label{eq:CRF_fusion}
\end{equation}
The two functions $F_{\text{segment}}(s_i,r,i)$ and $F_{\text{link}}(l_j,r,j)$ are state functions for the indicator variables $s_i$ and $l_j$, $r$ refers to the set of all regions obtained from the 2D region segmentation, and $n$ and $m$ are the total number of indicator variables $s_i$ and $l_j$. 
To ensure that any large segment from the region segmentations can compete equally against a set of smaller regions covering the same 2D area, both state functions take the size of the corresponding regions into account. In addition, links between regions are weighted according to the similarity of the linked regions, leading to the following definitions:
\begin{equation}
F_{\text{segment}}(s_i,r,i) = \text{size}(r_i)
\end{equation}
\begin{equation}
F_{\text{link}}(l_j,r,j) = \theta(r_{j,a},r_{j,b}) \cdot \big(\text{size}(r_{j,a}) + \text{size}(r_{j,b})\big).
\end{equation}
$r_{j,a}$ and $r_{j,b}$ are the two regions connected by the 3D link $l_j$ and $\theta(r_{j,a},r_{j,b})$ measures the similarity between two regions. 
In the original fusion formulation, Vazquez-Reina et al. defined $\theta$ in terms of the cross-correlation and displacement between the pair of segments that re connected by the link in question \cite{vazquez:11}. We instead define the region similarity $\theta$ in terms of the minimum relative overlap size of the two regions. This definition does not take texture similarity into account, but it is computationally faster than cross-correlation while providing equally good region similarity measurements for our EM data:

\begin{equation}
\theta(r_{j,a},r_{j,b}) = \min \Big( \text{overlap}(r_{j,a}, r_{j,b}), \text{overlap}(r_{j,b}, r_{j,a}) \Big)
\end{equation}
\begin{equation}
\text{overlap}(r_{j,a}, r_{j,b}) = \frac{|r_{j,a} \cap r_{j,b}|}{|r_{j,a}|}
\label{eq:fusion_links}
\end{equation}

By using the minimum of the relative overlap, $\theta$ is based on the relative overlap with respect to the larger region. This definition is useful, because if a large region is overlapped by two smaller regions by 40\% and 60\% respectively, we want the link to the region covering 60\% of the overlap to outweight the link to the region covering 40\%.

The compatibility function $\psi(s,l)$ in equation \ref{eq:CRF_fusion} is defined over the indicator variables and assigns zero-mass to configurations of the CRF that are unrealistic or undesirable given our domain knowledge of the problem. 
Specifically, we want each pixel of the segmentation to belong to no more than one neuron, thus we want to prevent overlapping of activated segments $s_i$ in the same section image: $\sum_{i \in o_k} s_i \leq 1$ for every set of overlapping segments $o_k$.
Furthermore, the selection of activated segments and links should yield good 3D continuity through the stack. We achieve this by making the selection of segments and links dependent on each other and rewarding the activation of segments that are connected by strong links: 

\begin{equation}
\left( \sum_{j \in \text{TOP}_i} l_j \right) \leq |\text{TOP}_i| \cdot s_i, 
\left( \sum_{j \in \text{BOT}_i} l_j \right) \leq |\text{BOT}_i| \cdot s_i
\end{equation}where $\text{TOP}_i$ and $\text{BOT}_i$ are the sets of activated links connecting to segment $s_i$ from the sections immediately above or below, respectively. The main idea is to make the activation of segments and links depend on each other. Whenever a link is activated, the corresponding connected segments have to be activated as well. Compared to the original fusion formulation, our constraint does allow for multiple links connecting to the same segment. This relaxed version allows for branching of segmented structures and thus can adequately model the geometry of neuronal cells. 

In our experiments, we noticed that if we allow segments to connect with any number of overlapping segments from adjacent sections, we have to be careful to not over-merge the segmentation. To prevent branching from connecting too many objects, we require links to only connect segments with significant overlap between sections. Links connecting sections that do not or only minimally overlap are pruned from the solution space. To obtain the maximum a posteriori (MAP) solution to the whole segmentation fusion problem (Eq. \ref{eq:CRF_fusion}), we solve the following binary linear programming problem:

\begin{equation}
\begin{aligned}
\operatorname{argmax}_{s,l} \sum_{i=1}^n & F_{\text{segment}}(s_i,r,i) + \sum_{j=1}^m F_{\text{link}}(l_j,r,j) \\
\text{s.t.} \; & s_i, l_j \in \{0,1\}, \\
& \psi(s,l) = 1.
\end{aligned}
\label{eq:blp_fusion}
\end{equation}

We solve this problem using a general-purpose binary linear programming solver \cite{ibm:cplex}.

%% file: results_fusion.tex
\subsection{Segmentation Fusion Evaluation}
There are two important aspects for the evaluation of segmentation fusion: the benefit of using multiple segmentations per section on the 2D segmentation, and the performance with respect to the 3D region grouping into geometrically consistent objects.
\begin{figure}%
\includegraphics[width=\columnwidth]{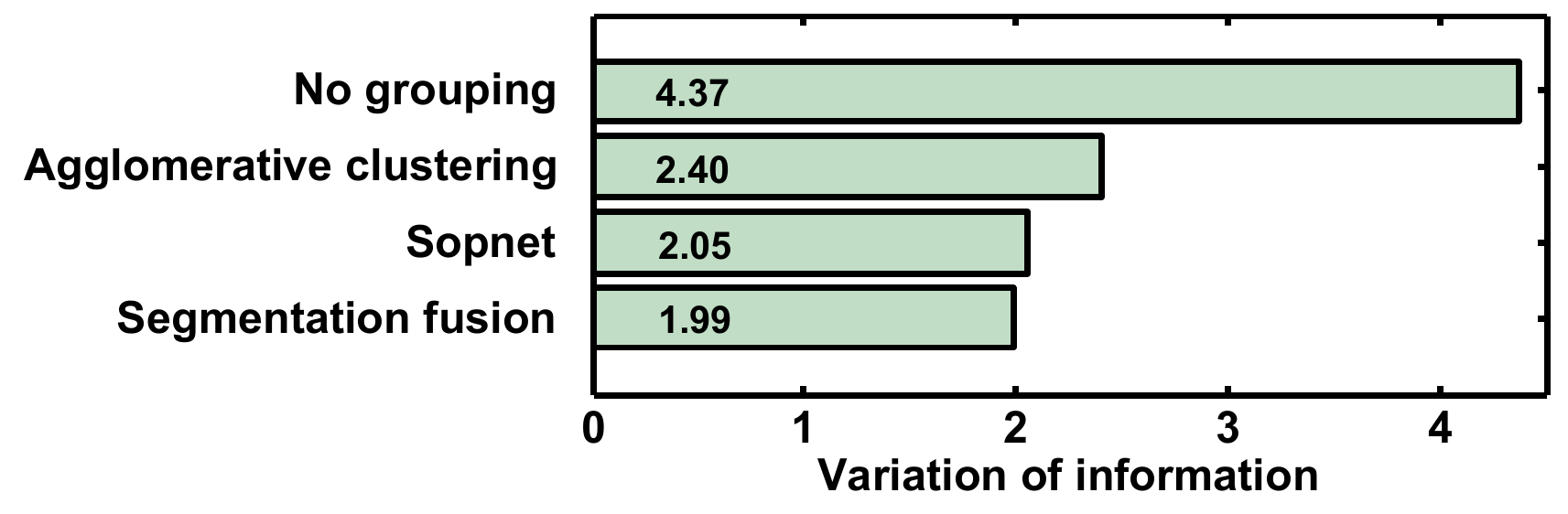}%
\caption{Evaluation of segmentation fusion compared to agglomerative clustering  and Sopnet \cite{Funke:12}. As a baseline we provide the variation of information score for ungrouped data. Lower scores correspond to better region groupings.}%
\label{fig:Results_fusion_regionGrouping}
\end{figure}
(see Figure \ref{fig:Results_fusion_regionGrouping}). To evaluate the 3D region grouping performance we compare against greedy agglomerative clustering \cite{kaynig:10b} and the Sopnet framework developed by Funke et al. \cite{Funke:12}. Segmentation fusion and the Sopnet framework both significantly outperform agglomerative clustering by finding the globally optimal grouping with respect to a large volume context. Sopnet and segmentation fusion obtain about the same quality of region groupings with segmentation fusion performing 0.06 better in terms of variation of information. Both approaches leverage multiple segmentation hypotheses per section and obtain the optimal grouping by solving an integer linear programming problem. The main difference is that within Sopnet a classifier is trained to score the similarity between regions, whereas segmentation fusion relies on region overlap and size alone (see section \ref{subsec:fusion}). It is possible that Sopnet could benefit from a training set larger than 75 images, but generating such a large training set would require a considerable effort of manual annotation. In addition, the feature extraction and classification to obtain the region similarity adds significant computational overhead to the region grouping. Thus, segmentation fusion is the better approach for our large-scale reconstruction effort. 

\begin{figure}%
\includegraphics[width=\columnwidth]{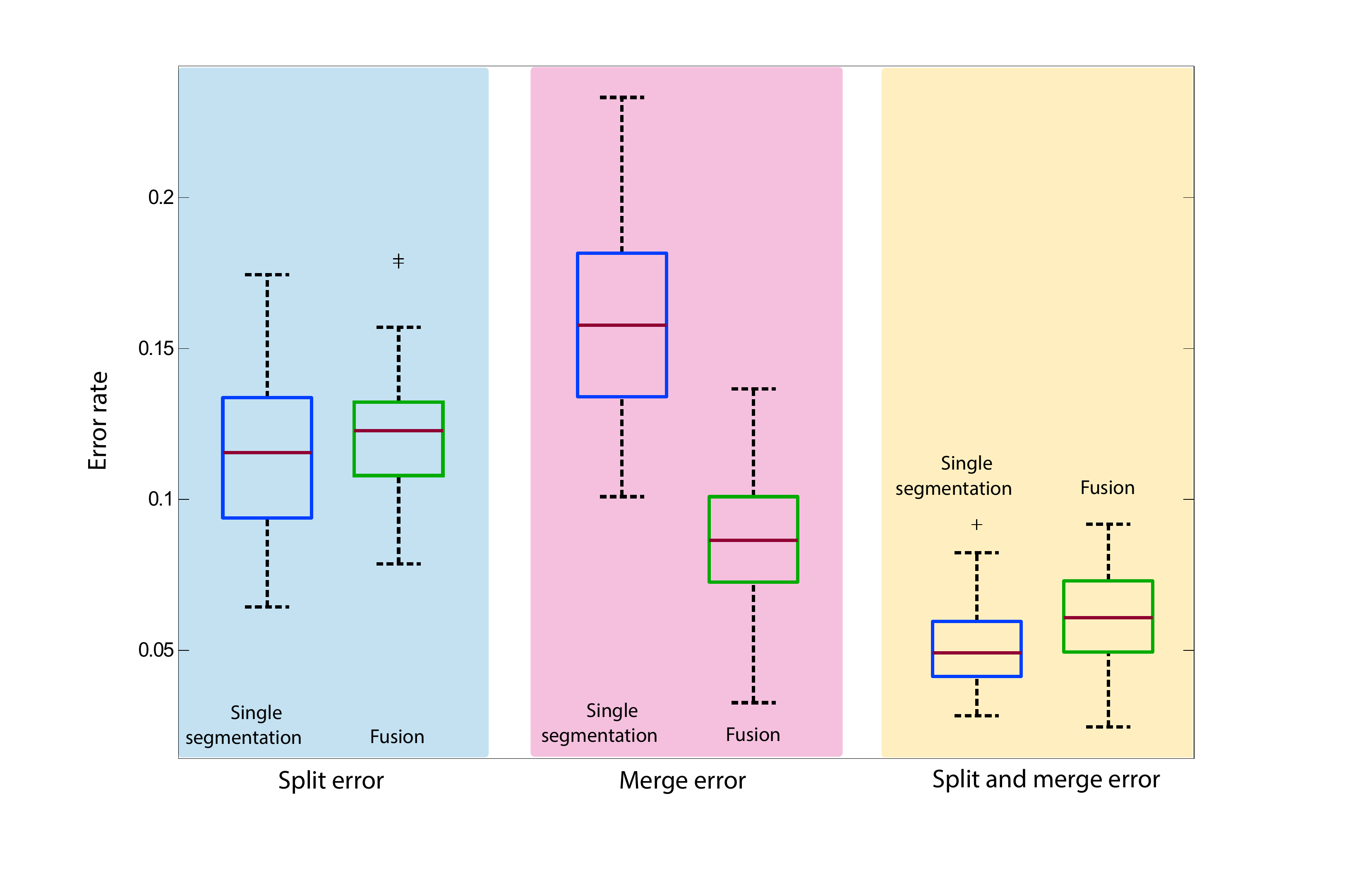}%
\caption{Improvement of split and merge errors in the 2D segmentations. Compared to the best single segmentation from the gap completion CRF output, fusion reduces the overall error rate by 5\% by leveraging multiple segmentations. The evaluation shows that this improvement is largely due to fewer merge errors.}%
\label{fig:Results_fusion_2d}%
\end{figure}

\begin{figure}%
\includegraphics[width=\columnwidth]{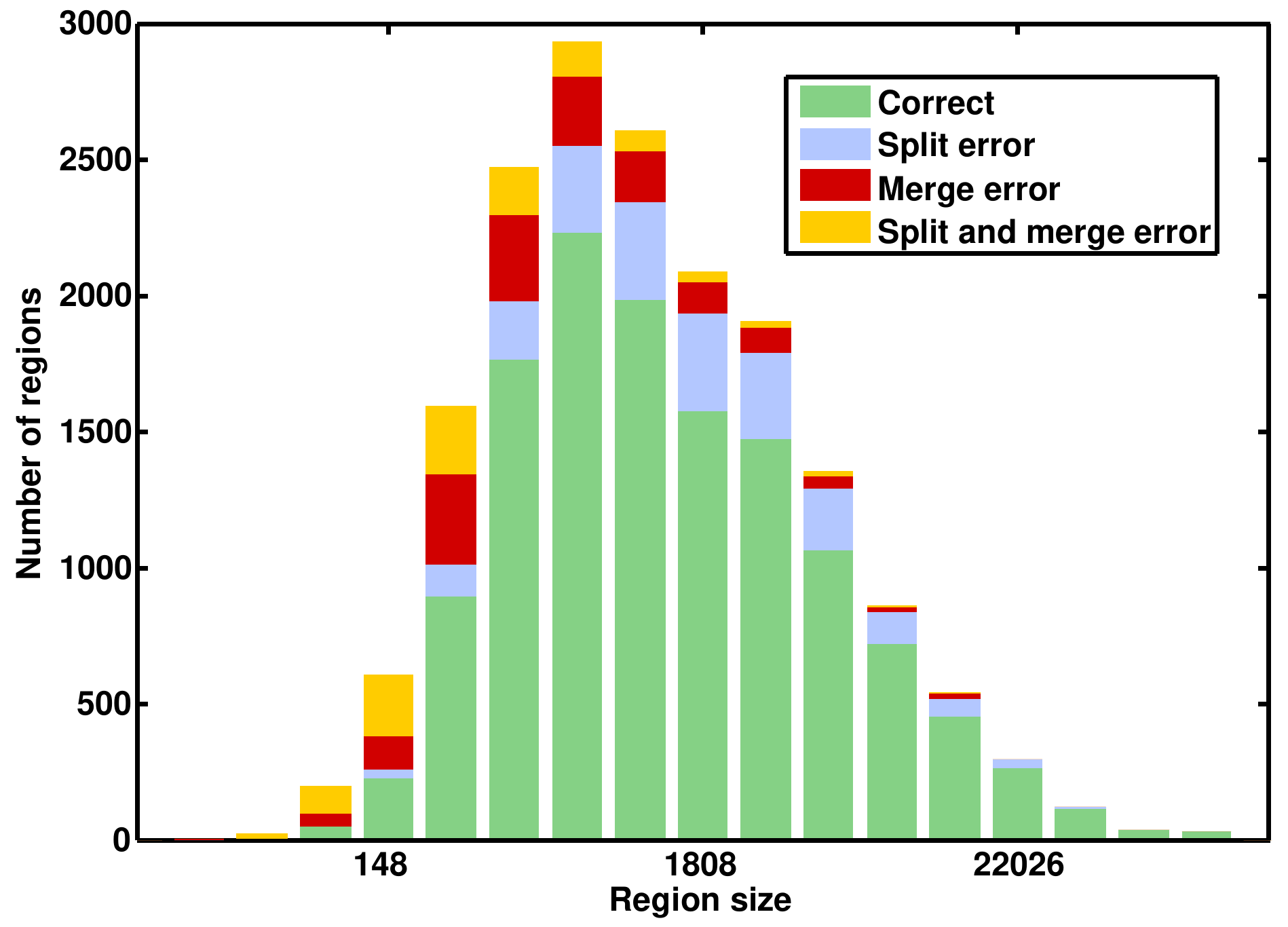}%
\caption{Histogram of error rates for different region sizes. While over 70\% of the regions are correctly segmented for most region sizes, smaller regions tend to be merged, whereas larger regions tend to be split. Overall large regions exhibit smaller error rates than small regions.}%
\label{fig:Results_fusion_2d_hist}%
\end{figure}

To gain more insight into the segmentation performance we evaluate 2D segmentations with respect to split and merge errors. Figure \ref{fig:Results_fusion_2d}
compares the segmentation fusion output with the best single 2D segmentation obtained by the gap completion CRF framework as described in Section \ref{subsec:regionsegeval}. Over the whole test set of 75 images, fusion gave a significant improvement in the overall segmentation performance, increasing the percentage of correctly segmented regions from 
$68\%(\pm 4\%)$ to $73\%(\pm 3\%)$ compared to the manual annotation. Figure \ref{fig:Results_fusion_2d} demonstrates that this improvement is mainly due to a correction of merge errors. While the split error rate is slightly incrased by 1\% the merge error rate is nearly halved, dropping from $15.8\%$ for the single segmentation to $8.6\%$ for the fusion result. Figure \ref{fig:Results_fusion_2d_hist}
shows the segmentation performance with respect to region sizes. In total, error rates are lower for the larger regions than for the small regions. It is also noteworthy that the dominant error type changes from mainly merge errors for smaller regions to split errors for larger regions. For the most typical region size the errors are nearly balanced. Because smaller regions belong to thin flexible neuronal processes they change more prominently between adjacent sections than large regions belonging to thick neuronal processes. Thus, it is challenging to pick the right region segments based on the overlap criterion as described in (Eq. \ref{eq:fusion_links}). This effect could be alleviated by estimating optical flow or a non-linear warping between adjacent sections, but these methods introduce a significant computational overhead to the segmentation, rendering them impractical for the large-scale reconstructions addressed by our work.

%% file: proofreading.tex
Manual proofreading is necessary in order to guarantee the correct topology of the neuron reconstruction. Figure 
\begin{figure}%
\centering
\includegraphics[width=.3\columnwidth]{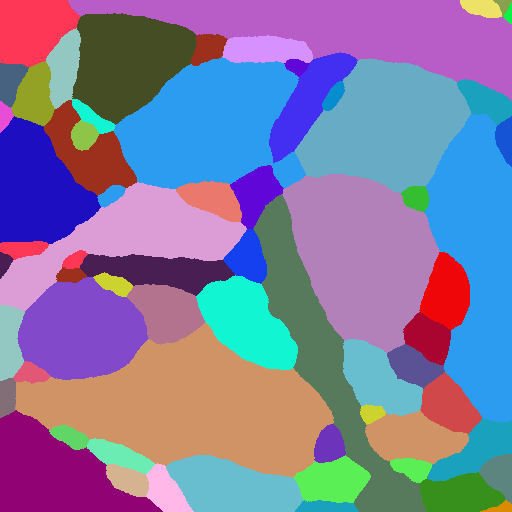}
\includegraphics[width=.3\columnwidth]{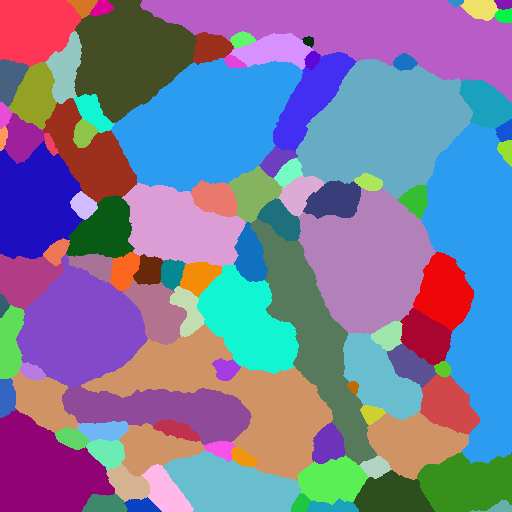}
\includegraphics[width=.3\columnwidth]{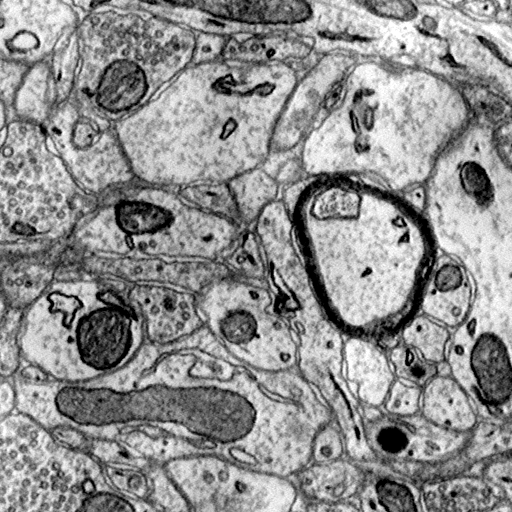}
\caption{Example region of an automatically segmented image (middle) compared to manual annotation (left). The original EM image is shown on the right.}%
\label{fig:example_section}%
\end{figure}
\ref{fig:example_section} shows an example segmentation of a 2D section compared to a manual annotation. While most regions are correctly segmented, some are split into several parts and need manual merging, while other regions span multiple objects and need to be manually split.

In order to minimize the user effort required to correct split and merge errors, we developed an interactive system called Mojo 
\begin{figure}[!t]
\centering
\includegraphics[width=\columnwidth]{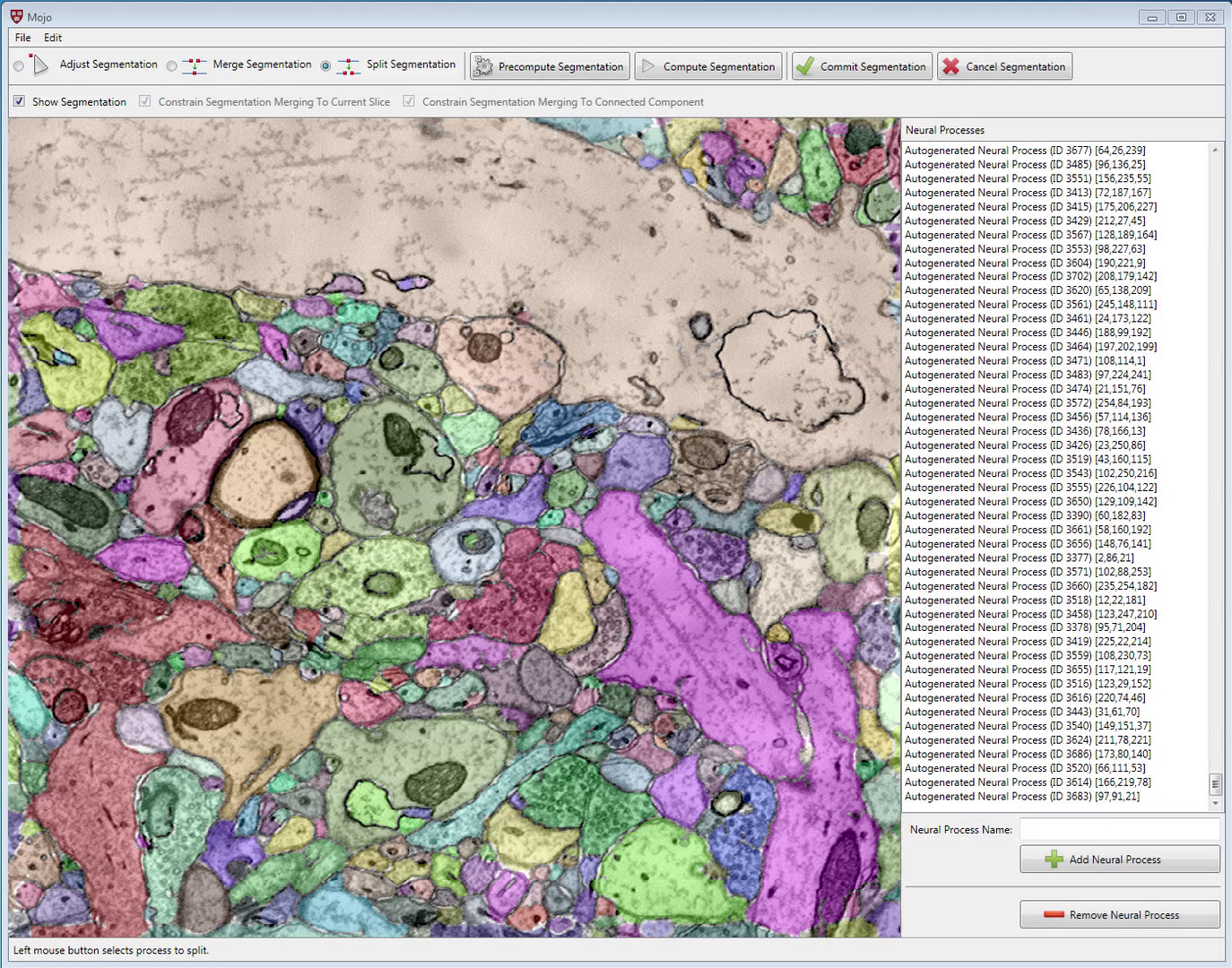}
\caption{Screenshot of our proofreading tool Mojo. The main area shows the segmentation as color-overlay on the original electron microscopy image. On the right side segmented processes can be selected by their name or ID. In the top menu the user can select split or merge error correction and if the correction should be applied in 3D or restricted to the currently shown section.}
\label{fig_mojo}
\end{figure}
(see Figure \ref{fig_mojo}). The proofreading workflow in Mojo follows a common workflow for manual neuron reconstruction: The user is presented with a 2D view of a 3D EM image stack that allows zooming, panning, and scrolling through the out-of-plane dimension. However, rather than requiring the user to draw each neuron cross-section by hand, Mojo presents the user with an interactive color overlay of the automatic reconstruction. Mojo provides two distinct interaction modes for correcting split errors and merge errors, respectively.

Correcting a split error requires the user to merge objects. To correct a split error in Mojo, the user clicks on a \emph{source} object, and then clicks on one or more \emph{target} objects. Mojo is responsible for re-labeling the target objects with the label from the source object. Our collaborators have found the following merge modes to be useful during proofreading:

\begin{itemize}
\item {\bf Global.} In this mode, Mojo re-labels all target voxels with the source label, regardless of the target voxel's location in the image stack.
\item {\bf Local-3D.} In this mode, Mojo re-labels only those target voxels that belong to the 3D connected component selected by the user.
\item {\bf Local-2D.} In this mode, Mojo re-labels only those target voxels that belong to the 2D connected component selected by the user.
\end{itemize}

Correcting a merge error requires the user to split an object into multiple sub-objects. The user begins by clicking on the object to be split. The user then roughly scribbles with a uniquely colored brush on each distinct sub-object within the object to be split. We use the interactive segmentation method of Roberts et al. \cite{roberts:11} to segment each sub-object. We chose to implement this segmentation method in Mojo because it produces highly accurate segmentations with minimal user effort, and provides the user with real-time feedback on the resulting segmentation while the user is scribbling. During each split operation, we constrain the user scribbles and the resulting sub-objects to be entirely contained within the object to be split. This allows the user to more easily segment each sub-object without disturbing neighboring objects.

After proofreading with Mojo, it is not guaranteed that all objects with the same label are topologically connected. When designing Mojo, we chose not to enforce such topological constraints on split and merge operations to allow for greater user flexibility. It is often the case that a user arrives at meaningful error corrections by \emph{composing} several split and merge operations (e.g., splitting off part of an object, then merging it with a neighboring object, and so on), eventually arriving at a single fully error-corrected object. In such cases, \emph{individual} split and merge operations may temporarily break an object's topology, even though the user's \emph{composition} of split and merge operations preserves it. Indeed, enforcing topological connectedness constraints in such cases would create extra work for the user. For example, if Mojo automatically forced an object to have two different labels when the object was temporarily broken into two pieces, the user would often have to undo these automatic changes as part of a higher-level composition of split and merge operations. By choosing not to enforce topological connectedness constraints on individual split and merge operations, we allow the user to more conveniently express a wider class of meaningful error corrections. However, this flexibility comes at the expense of relying on users to enforce any topological constraints required in their intended biological analyses.
 
Currently we are expanding Mojo to enable proofreading of data sets of arbitrary size. Mojo was used successfully to browse the large-scale reconstruction data described in this paper. To cope with such large data sets, fixed size image tiles  (each $512 \times 512$ pixels) were pre-computed at multiple zoom levels. At runtime, only tiles that are currently visible are loaded. The zoom level of the tiles loaded is determined by the user's current zoom level. This design ensures that Mojo can operate on a fixed memory budget of less than 1GB, while still being able to load data sets that are much larger than the available memory of the workstation on which it is running. A non-trivial task is merging of large volumetric objects in GB data sets in real-time. Relabeling of all object ids after an error correction can become a bottleneck in the interactive workflow. Currently we use a lookup table to restrict the relabeling to tiles containing the structure of interest. In practice this approach is generally fast enough, as most proofreading operations involve merging of smaller structures into a large object or splitting "`slabs"' from an object, caused by a merge error in a single section. Further speedup for splitting and merging large objects is part of our future research.

%% file: implementation_and_scalability.tex
Our pipeline has been designed to efficiently scale to large data sets in the GB-TB range. In the following sections we describe the run-time performance and scalability of the current implementation. All evaluations concerning run time are given with respect to the current MATLAB implementation and include computational and i/o overhead to facilitate restarting of jobs on a computer cluster. 

\paragraph{Membrane classification and segmentation generation} The 2D segmentation generation part of the pipeline is computed by a single job per image, leading to a total of 1000 jobs for the entire data volume. Each job performs the feature extraction, membrane classification and CRF segmentation with a typical runtime of 9-13 hours per job. 

\paragraph{Segmentation fusion} For this step, we divide the data into 3200 sub-volumes. For each volume one job computes the adjacency matrix between regions and the corresponding weights, and then solves the fusion problem to obtain the optimal region grouping for this sub-volume. The average runtime is 2-3 hours per job. 

\paragraph{Pairwise matching} In this step, sub-volumes are joined to form a consistently labeled volume. Every pair of adjacent sub-volumes (in x, y and z directions) is considered independently and in parallel. Winning groups of segments from the fusion step are merged or split based on the proportion of overlapping voxels inside the overlap region. Merge operations link segments to form a single object. Split operations assign all 2D segments in the overlap to just one group, creating two non-overlapping objects. This ensures a consistent labeling of voxels between adjacent sub-volumes. Runtime for each pair of sub-volumes is about 5-6 minutes, and 8120 sub-volume pairs are required for the full volume.

\paragraph{Global merging and output generation} A single global consistency step is required to link objects over multiple sub-volume pairs. Results from the fusion and pairwise matching steps are loaded and each connected component in the volume is assigned a unique label. This step is currently performed by a single job and takes 1-2 hours to load the data, compute the result, and write the output files.

\paragraph{Proofreading with Mojo} The implementation of our proofreading tool is optimized and implemented to run on a GPU \cite{roberts:11}. The main limitation for this step of the pipeline is not the computational performance, but the human proofreader. Parallel proofreading of the same image volume requires locking mechanisms or tedious resolving of conflicts between different annotators. A possible solution is to assign the tiled sub-volumes from our pipeline output in parallel to different proofreaders and only merge the different cubes after the proofreading step, using pairwise matching.

%% file: results_largeDataSet.tex
\section{Large-scale reconstruction results}\label{sec:large_scale_reconstructions}
We successfully used our pipeline to reconstruct neuronal processes in a $\mathrm{27,000 \; \mu m^3}$ volume of brain tissue. The following reconstruction results were obtained automatically, without any manual proofreading.
Figure 
\begin{figure*}%
\centering
\includegraphics[trim = 25cm 0mm 0mm 0mm, clip=true, width=\textwidth]{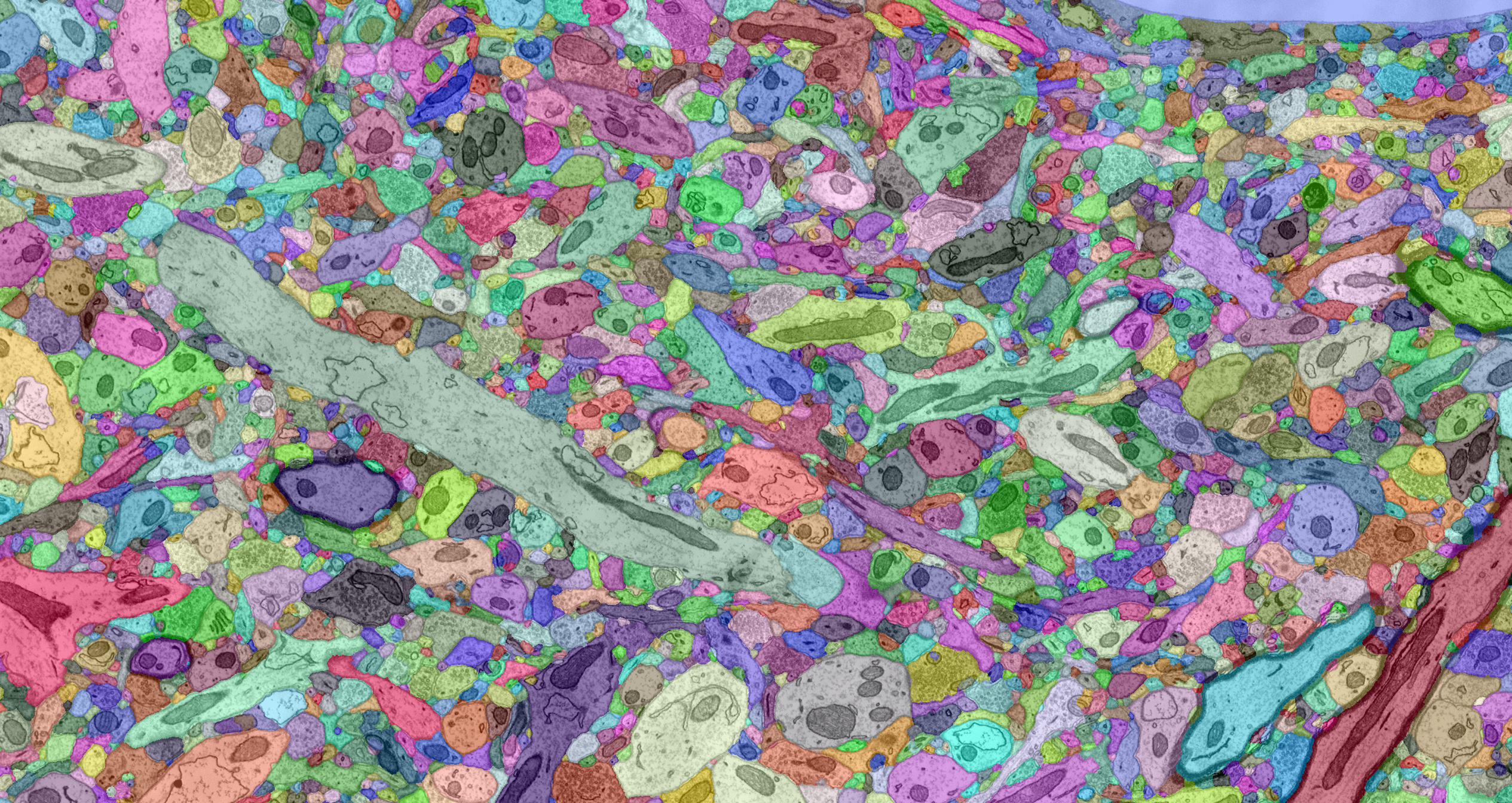}%
\caption{Example segmentation of a region of interest from the reconstructed EM volume. The image shows an overlay of the segmentation in color on the gray value EM image. This result is prior to manual proofreading.}%
\label{fig:Results_largeScale_stainedGlass}%
\end{figure*}
\ref{fig:Results_largeScale_stainedGlass} depicts an example segmentation of 2D image from our reconstructed volume. 

Figure     
\begin{figure}%
\centering
\includegraphics[width=\columnwidth]{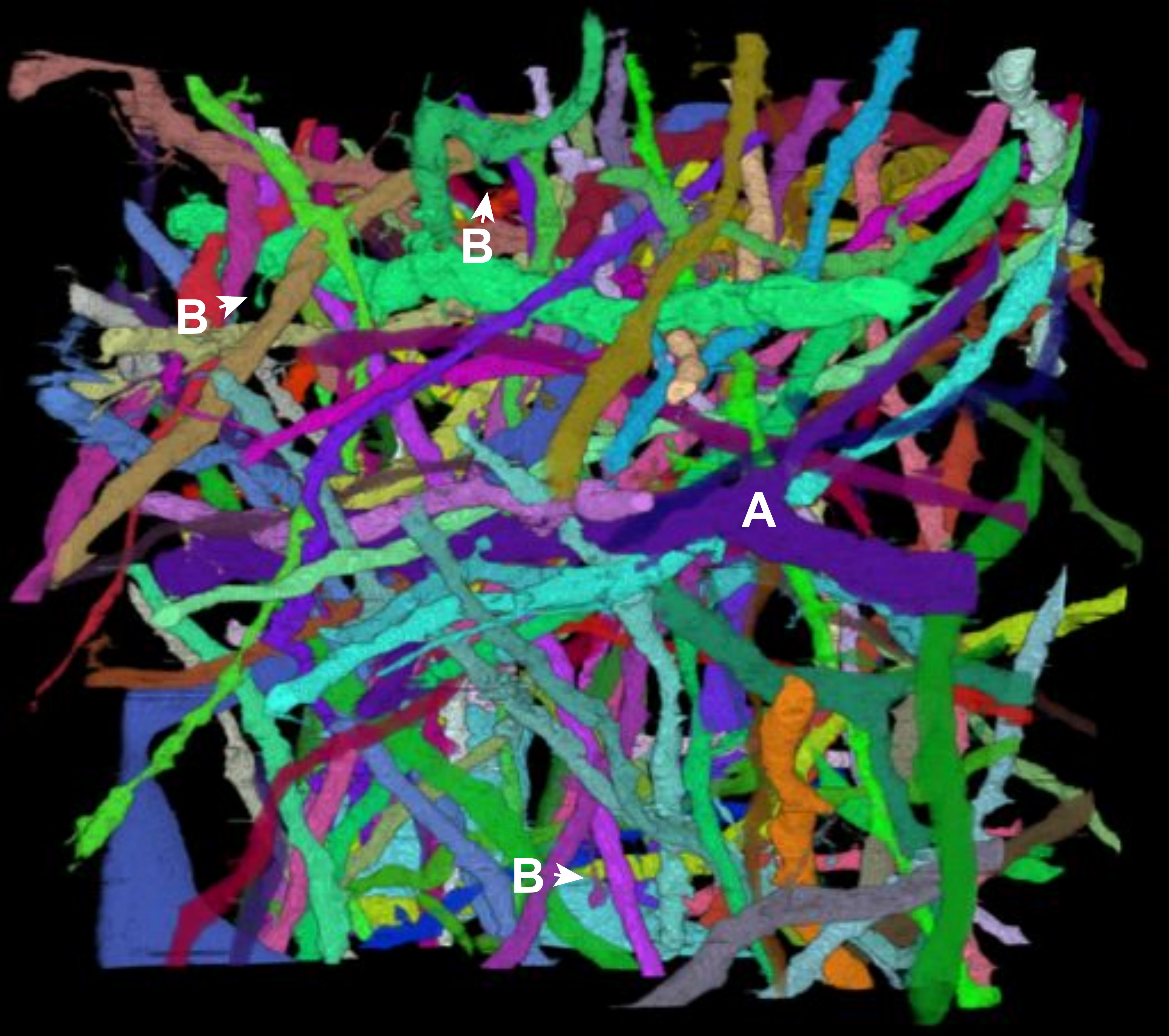}%
\caption{Automatic large-scale reconstruction of 93 objects from the whole data volume without any manual proofreading. The reconstructions contain long-range tracings across the whole volume running orthogonal across all 1000 sections as well as longitudinal to the cutting plane. A correctly reconstructed branching structure is marked with (A). (B) marks automatically reconstructed spine necks of approximately $\mathrm{30 \; nm}$ in diameter.}%
\label{fig:Results_largeScale_side_to_side}%
\end{figure}
\ref{fig:Results_largeScale_side_to_side} shows a subset of the reconstructed processes. As the chance of introducing an error to the reconstruction of an object grows exponentially with the object length it is important to evaluate the long range performance of an automatic reconstruction. Therefore the visualization in Figure \ref{fig:Results_largeScale_side_to_side} only includes processes that are traced from one face of the volume to another and that do not show obvious errors in the reconstructed 3D geometry. In total 93 objects satisfied these criteria. Note that although our data set is anisotropic, the reconstruction contains processes that run orthogonal as well as horizontal across the volume. The annotations mark an example of a correctly reconstructed branching structure (A) as well as several reconstructed spine necks (B). 

Spine necks are important parts of neuronal processes in mammalian tissue, as their spine head normally ends in a synapse forming a connection to another neuronal process. They also form the thinnest parts of a neuronal process and can have a diameter of only $\mathrm{25 \; nm}$. Thus they can be hard to distinguish from extracellular space between cells and sometimes are also missed by human expert annotators. Figure 
\begin{figure}%
\centering
\includegraphics[width=.5\columnwidth]{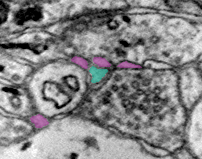}%
\caption{Example image with an annotated spine neck (green) and multiple annotated extracellular space regions (purple). The small diameter of spine necks makes the automatic 3D reconstruction challenging.}%
\label{fig:spine_neck_vs_extracellular_space}%
\end{figure}
\ref{fig:spine_neck_vs_extracellular_space} shows an example image with annotations for a spine neck region and several regions corresponding to extracellular space. The small diameter of spine necks renders their automatic 3D reconstruction challenging. Differentiation of spine neck regions and extracellular space is often only possible by taking the broader 3D context into account. To gain more insight into the quality of our spine neck reconstructions we used a manually annotated part of a dendrite and cut out the corresponding area from the automatic reconstruction. Figure
\begin{figure}%
\includegraphics[width=\columnwidth]{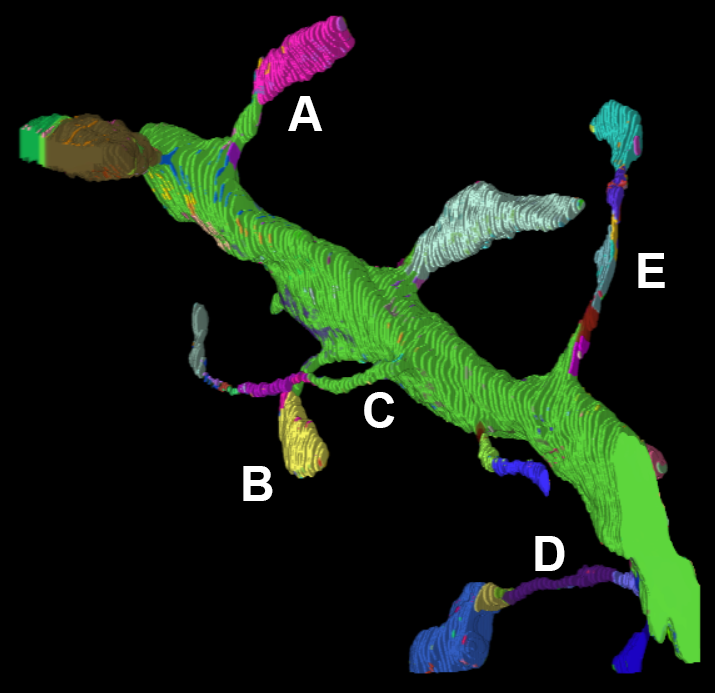}%
\caption{Fragmentation of a automatically reconstructed dendrite with respect to a manual reconstruction. The different colors correspond to different objects in the automatic segmentations. While manual proofreading is necessary to obtain the correct reconstruction, the automatic result includes spine necks traced over several sections. (A) and (B) only require merging of the identified spine neck with the spine head, whereas (C) and (D) exhibit further split errors, but still contain large continuous segments. (E) is an example of a fragmented spine neck, running longitudinal to the cutting plane.}%
\label{fig:gt_crop_out_spine_necks}%
\end{figure}
\ref{fig:gt_crop_out_spine_necks} depicts the result. While none of the spines are correctly reconstructed automatically, proof reading the fragmentation of spines (A) and (B) is reduced to correcting one split error. The spines (C) and (D) are more fragmented than (A) and (B), but contain tracings over several sections. (E) is heavily fragmented as it runs longitudinal across the sections and thus is harder to trace than orthogonal spine necks. 

Region grouping with branching is essential not only to account for branching neuronal processes, but also to reconstruct spine necks along a dendrite. However, it can also lead to merge errors in the resulting segmentation.
Figure 
\begin{figure}%
\includegraphics[width=\columnwidth]{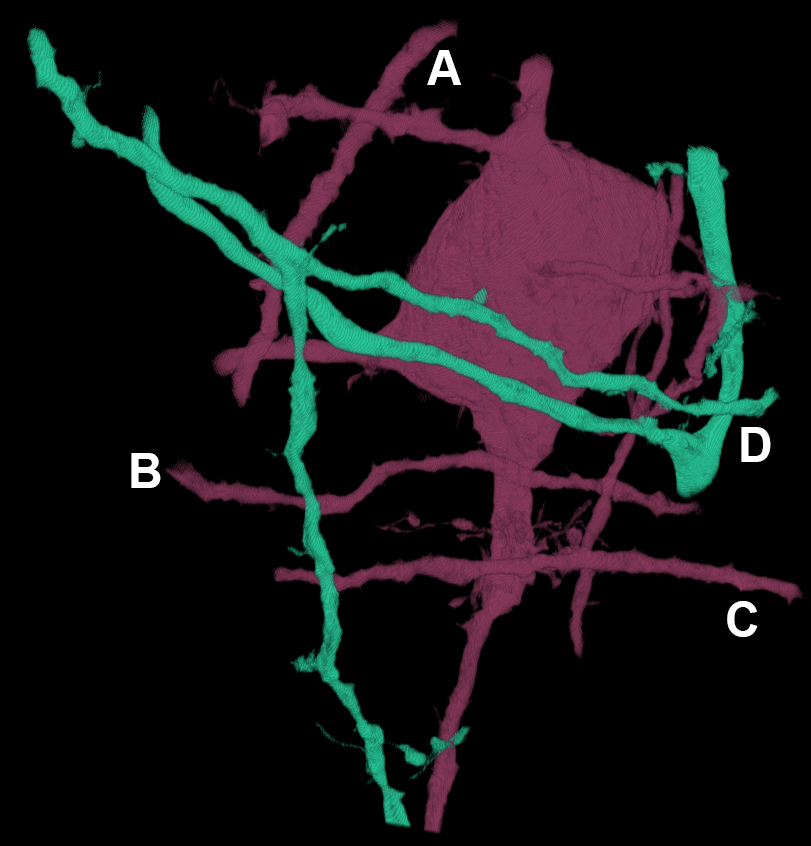}%
\caption{Examples of large-scale merge errors. The large red cell body is merged with neuronal processes marked (A), (B), and (C). The green object contains two branching structures, which are erroneously merged at location (D).}%
\label{fig:Large_scale_merge_errors}%
\end{figure}
\ref{fig:Large_scale_merge_errors} shows examples of long range merge errors. The reconstruction shown contains a correctly segmented cell body (red) including correct branches. The neuronal processes marked (A), (B), and (C) should be separate objects. The green structure is a merge of two neuronal processes. While both processes contain a correctly identified branching point, they are erroneously merged at location (D). These long range merge errors are easy to detect for a human proof reader by looking at the 3D geometry of the reconstructed objects. Automatic identification and correction of the long range 3D geometry is part of our future research.

%% file: conclusion.tex
In this paper we address the automatic reconstruction of neuronal processes at $\mathrm{nm}$ resolution for large-scale data sets. We demonstrate state-of-the art performance of our pipeline with respect to automatic dense reconstruction of neuronal tissue, and also for long range reconstructions covering neuronal processes over many $\mathrm{\mu m}$. The workflow is designed to minimize manual effort and to be easy parallelizable on computer clusters and GPUs, with most steps scaling linearly with the number of processors. 

Future work concentrates on improving the performance of our pipeline, as well as facilitating the proofreading further. We are currently focusing our efforts on improving the runtime of the pipeline by optimizing code and removing MATLAB dependencies. We are also working on new algorithms to improve the overall segmentation performance. With respect to proofreading, we are working on a web-based Mojo version that allows for collaborative proof reading of large data volumes. 